\documentclass[twocolumn,pra,aps,superscriptaddress,nofootinbib]{revtex4-2}
\usepackage[english]{babel}
\usepackage[utf8]{inputenc}

\def\Z{\mathbb{Z}}

\usepackage{graphicx}
\usepackage{amsmath,amsthm}
\usepackage{tikz}
\usepackage{amsfonts}
\usepackage{amssymb}
\usepackage{multirow}
\usepackage{mathrsfs}
\usepackage{makeidx}
\usepackage{amsmath}
\usepackage{hyperref}
\usepackage{placeins}

\def\({\left(}
\def\){\right)}
\def\<{\left\langle}
\def\>{\right\rangle}
\def\beq{\begin{equation}}
\def\eeq{\end{equation}}

\def\x{\mathbf{x}}

\def\y{\mathbf{y}}

\def\KS{\text{KS}}

\def\max{\text{max}}
\def\opt{\text{opt}}

\begin{document}

\title{Emergent correlations in the selected link-times along optimal
  paths}

\author{Iván Álvarez Domenech}
\affiliation{Dto. Física Matemática y de Fluidos, Universidad Nacional
  de Educación a Distancia (UNED), Madrid (Spain)}

\author{Javier Rodríguez-Laguna}
\affiliation{Dto. Física Fundamental, Universidad Nacional de
  Educación a Distancia (UNED), Madrid (Spain)}

\author{Pedro Córdoba-Torres}
\affiliation{Dto. Física Matemática y de Fluidos, Universidad Nacional
  de Educación a Distancia (UNED), Madrid (Spain)}

\author{Silvia N. Santalla}
\affiliation{Dto. Física \& GISC, Universidad Carlos III de Madrid,
  Leganés (Spain)}

\begin{abstract}
In the context of first-passage percolation (FPP), we investigate the
statistical properties of the {\em selected link-times} (SLTs) --the
random link times comprising the optimal paths (or geodesics)
connecting two given points. We focus on weakly disordered square
lattices, whose geodesics are known to fall under the
Kardar-Parisi-Zhang (KPZ) universality class. Our analysis reveals
universal power-law decays with the end-to-end distance for both the average
and standard deviation of the SLTs, along with an intricate pattern of
long-range correlations, whose scaling exponents are directly linked
to KPZ universality. Crucially, the SLT distributions for diagonal and
axial paths exhibit significant differences, which we trace back to
the distinct directed and undirected nature, respectively, of the underlying geodesics.
Moreover, we demonstrate that the SLT distribution violates the
conditions of the central limit theorem. Instead, SLT sums follow the
Tracy-Widom distribution characteristic of the KPZ class, which we
associate with evidence for the emergence of high-order long-range
correlations in the ensemble.
\end{abstract}

\date{January 24, 2026}

\maketitle

\section{Introduction}

Selection and conditioning often induce correlations in otherwise
independent and identically distributed (iid) random variables. Simple
examples include the Brownian bridge, i.e. random walks constrained to
start and end at the same points \cite{Karatzas.88}, and order
statistics, i.e. sorted iid variates \cite{David.03}. In both cases,
emergent long-range correlations among random variables appear as a
result of either a global constraint or a sorting procedure. Another
counterintuitive effect is known as {\em collider bias} or Berkson's
paradox, i.e. selecting individuals above a certain threshold for the
sum of different uncorrelated traits induces a spurious correlation
between them \cite{Berkson.46,Pearl.18,DeRon.21}. This explains the
existence of unexpected negative correlations between luminosity and
distance in flux-limited astronomical catalogs (Malmquist bias)
\cite{Butkevich.05}, or the apparent ties between independent
lifetimes in risk assessment \cite{Prentice.78}. Finally, in
quantitative genetics, evolutionary selection of a trait which is the
sum of several effects associated to independent loci tends to reduce
its variance through the emergence of negative correlations among
them, which is known as Bulmer effect \cite{Bulmer.71}. Our starting
hypothesis is that such emergent correlations must also appear among
the link-times which are actually selected by geodesics in random
media --also known as optimal paths--, even though the whole ensemble
of link-times remains uncorrelated.

The characterization of {\em optimal paths} in disordered networks
\cite{Havlin.2005,Bhattacharyya.2005}, also known as {\em
  first-passage percolation} (FPP) \cite{Hammersley.1965,
  Auffinger.2017}, is deeply connected to other relevant systems in
statistical mechanics, such as geodesics in random metrics
\cite{Santalla.2015,Santalla.2017}, minimum spanning trees
\cite{Dobrin.2001,VanMieghem.2005,VanMieghem.2005b,Wu.2006b,Wang.2008},
directed polymers in random media \cite{HalpinHealy.1995}, and
percolation models \cite{Lopez.2012}. Moreover, it is also crucial for
several applications involving random media, including transport in
communication networks
\cite{Wang.2008,VanMieghem.2005,VanMieghem.2005b,Wu.2006b,Fortz.2002,Mahajan.2002},
fluid or current flow
\cite{Barabasi.1996,Cieplak.1996,Porto.1997,Andrade.2000,Lopez.2005,Wu.2005,Li.2007},
magnetotransport \cite{Strelniker.2004,Strelniker.2006}, epidemic
spreading \cite{Tolic.2018}, and fracture \cite{Andrade.2009}.
Specifically, we will consider the FPP model on square lattices for
which the crossing times associated to the edges, or {\em link times}
(LTs), are weakly disordered iid random variables
\cite{Villarrubia.2024}. Interestingly, the statistical properties of
geodesics between two given points fall within the 1+1D
Kardar-Parisi-Zhang (KPZ) universality class
\cite{Cordoba.2018,Alvarez.2024}, which first appeared in the analysis
of growing interfaces \cite{Kardar.1986,Barabasi.1995}, but which has
since found applications in numerous areas of physics and mathematics
\cite{HalpinHealy.2015}. Indeed, the fluctuations in the arrival time
$T$ of a geodesic spanning a Euclidean distance $d$ are known to scale
as $\sigma_T \sim d^\beta$, where $\beta=1/3$ is the growth exponent
of KPZ, while its average lateral deviation from the straight line
scales as $h \sim d^{1/z}$, where $z=3/2$ is the dynamic exponent
\cite{Cordoba.2018}. Moreover, systems within the KPZ class tend to
present universal local fluctuations, corresponding to one of the
probabilty distributions in the Tracy-Widom (TW) family
\cite{HalpinHealy.2015}. We have recently shown \cite{Alvarez.2024}
that the local fluctuations of the isochrones follow the TW
distribution associated with the Gaussian unitary ensemble (TW-GUE),
and there is evidence that the latter could also govern the arrival
time fluctuations \cite{Cordoba.2018}. Finally, let us stress that the
geodesic length $\ell$ is proportional to the end-to-end distance $d$
in all the considered cases.

In this work we aim to characterize the ensemble of {\em selected
  link-times} (SLTs), i.e., the LTs which are actually chosen by
optimal paths spanning different Euclidean distances, in the
weak disorder regime. Naturally, we expect the SLTs to take lower
values than those of the whole ensemble of LTs, but {\em how much
  lower?} The reduction is expected to be smaller near the geodesic
ends --because of the small number of choices-- than in the geodesic
bulk. Crucially, the sum of the SLTs along a geodesic corresponds to
the total arrival time, which presents non-Gaussian statistics.
Therefore, the SLTs must violate the hypotheses of the central limit
theorem (CLT) \cite{Feller.68,Ibragimov.71,Billingsley.95}. Such
violation, as we will see, takes place through the emergence of
long-range non-Wick correlations of higher order
\cite{Peccati.11,Pipiras.17}, which we will associate to KPZ scaling
using an Ansatz inspired in 2D conformal field theory (CFT)
\cite{DiFrancesco.96}. This way, we will make contact with the area of
     {\em full-counting statistics} (FCS)
     \cite{Camia.16,Fewster.18,Anthony.19}, in which integrals of a
     certain physical density (energy, magnetization, etc.) subject to
     random fluctuations with high-order long-range correlations are
     shown to evade the constraints of the CLT.

This article is organized as follows. Section \ref{sec:model}
describes the basic properties of our physical model. The global
probability distribution of the SLTs is discussed in Sec.
\ref{sec:global}, while Sec. \ref{sec:local} characterizes the
dependence on the position along the geodesic and the link direction.
Section \ref{sec:correlations} discusses the emergent correlations among
the SLTs, both for two-point and higher-order correlators, which are
required to violate the conditions of the CLT and give rise to the TW
distribution of the arrival times. The article ends with a summary of
our conclusions and proposals for further work.


\section{Model}
\label{sec:model}

Let us consider the square lattice $\Z^2$, whose edge set is
denoted by ${\cal E}$. We associate to each edge $e\in {\cal E}$ an LT
$t_e>0$, which are. The set of LTs are iid random variables with common probability
distribution $F(t)$ ($F(0)=0$), probability density function (pdf)
$f(t)$, mean $\tau$, and deviation $\sigma$. A path $\gamma$ of length
$\ell=|\gamma|$ is defined as a sequence of edges
$\gamma=\{e_i\}_{i=1}^\ell$ such that $e_i$ and $e_{i+1}$ share only
one endpoint. The crossing time associated to the path is given by
$T(\gamma)=\sum_{e\in\gamma}t_e$. Now, given two nodes $\x,\y\in
\Z^2$, we may consider the path joining them presenting the minimum
crossing time, that is, the path $\gamma_\opt$ such that

\beq
T(\gamma_\opt)=\min_{\gamma\in\Gamma(\x,\y)}T(\gamma),
\label{eq.1}
\eeq
where $\Gamma(\x,\y)$ is the set of paths joining $\x$ with $\y$. This
path $\gamma_\opt$ is called the {\em geodesic} or {\em optimal path} between $\x$ and $\y$,
and the minimum crossing time associated is called the {\em arrival
  time} between the two nodes, denoted as $T(\x,\y)\equiv
T(\gamma_\opt)$. For notational convenience, we will set the initial
node of all our geodesics at the origin of coordinates, $\x=0$, and
denote $T(\x)\equiv T(0,\x)$. We may define now the ball at time $t>0$
around the origin, $B(t)$, as

\beq
B(t)=\{\x\in\Z^2\: :\: T(\x)\leq t\}.
\label{eq.2}
\eeq
The boundary of $B(t)$, $\partial B(t)$, will be called the {\em
  isochrone} correponding to time $t$.

When the disorder is weak, the coefficient of variation of the LT
distribution, CV$\equiv\sigma/\tau$, plays a central role in the
statistical properties of our system \cite{Cordoba.2018}. For CV$\ll
1$, the isochrones converge towards a diamondlike limit shape that reflects the geometry of the underlyinng lattice
\cite{Alvarez.2024}. Yet, as the CV increases, the diamond transforms
continuously into a circumference, and we observe isotropic growth. As
it was mentioned earlier, the deviation of the arrival times between
points separated a large distance $d$ grows as $\sigma_T(d)\sim
d^\beta$, where $\beta=1/3$ is the growth exponent of KPZ. Along a
diagonal, this asymptotic behavior takes place for all $d$, but along
an axis, the deviation initially grows as $\sigma_T(d)\sim d^{1/2}$, up
to a certain crossover distance, $d_c\equiv \tau^2/(3\sigma^2)$. The
difference between both cases is that the geodesic along the axis is
non-degenerate in the homogeneous case, i.e. when all the LTs are
equal \cite{Cordoba.2018}. Moreover, the average lateral deviation
between the geodesic and the straight line joining the endpoints,
$h$, scales as $h\sim d^{1/z}$ for large $d$, where $z=3/2$ is the
dynamic exponent of KPZ. Yet, if the points lie along an axis, the
lateral deviation is negligible for $d\ll d_c$ \cite{Cordoba.2018}.

\medskip

In this work we have characterized the statistical properties of the SLTs in geodesics both along the axial and diagonal directions on a $(2L+1)\times
(2L+1)$ square lattice with $L=1005$. Two different LT distribution
families have been employed to that end. On the one hand, the uniform
distribution on the interval $[a,b]$, for which the maximal attainable
value of the CV is $1/\sqrt{3}\approx 0.57$, since $a\geq 0$ necessarily.
On the other hand, we have used the Weibull distribution, given by the
probability density function

\begin{equation}
  f(t)=\frac{k}{\lambda}\(\frac{t}{\lambda}\)^{k-1}
  \exp\(-(t/\lambda)^k\),
\end{equation}
which allows for any positive value of the CV.

In all our simulations, we fix the mean value of the LT distribution,
$\tau=5$, and choose different values of the CV in order to survey the
different observables. Thus, we use the notation U(CV) and Wei(CV)
respectively for the uniform and Weibull distributions with parameter
CV. Also, to perform our statistical analysis we employ
$N_s=2\cdot 10^4$ different noise realizations. Finally, the maximum end-to-end distance considered here will be $d_{\max}=1000$ for the axis and $d_{\max}=1000\sqrt{2}$ for the diagonal.


\section{Global SLT distribution}
\label{sec:global}

Geodesics tend to pass through links with low crossing times, i.e.
the SLTs tend to be lower than the original LTs, and in this section
we characterize the difference between both distributions. We will
denote by $\hat f(t)$ the {\em global} pdf for the SLTs along the
different geodesics, and we will extend the notation to the global
probability function $\hat F(t)$, and to the mean and standard deviation, $\hat\tau$ and $\hat\sigma$, respectively.

\subsection{Mean and deviation of the global SLT distribution}

\begin{figure}
\includegraphics[width=7cm]{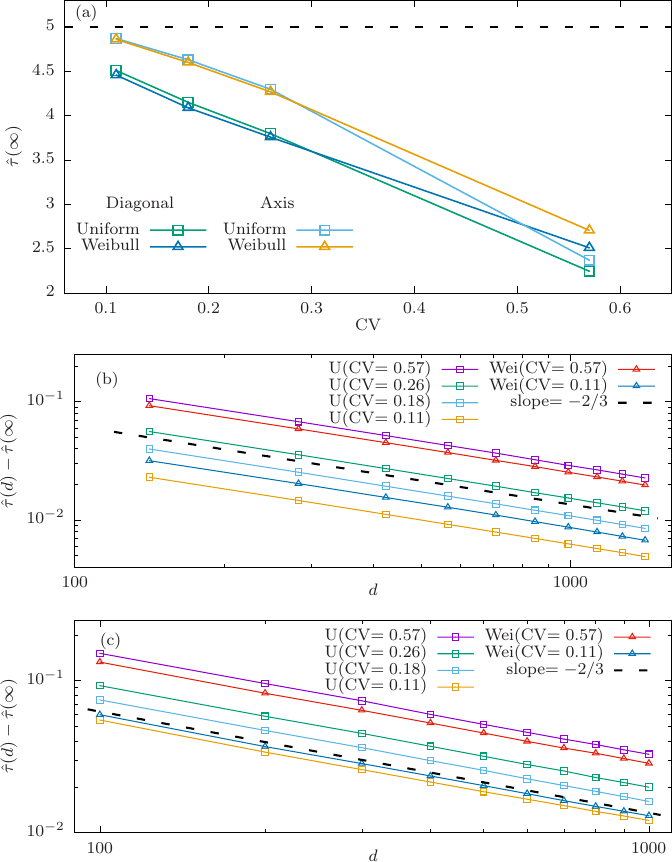}
\caption{(a) Asymptotic value of the global SLT mean as as function of the CV for the two distributions and the two lattice directions. The broken line indicates the LT mean. (b) Convergence of the SLT
  averages towards their asymptotic values for geodesics along the
  diagonal, as we increase the end-to-end distance $d$. (c) Same data,
  along the axis. The broken line in both panels represents power-law behavior with exponent $-1/z$.}
\label{fig:global_tau}
\end{figure}

In this section we will consider the dependence of the first two moments of the global SLT distribution on the end-to-end distance $d$, i.e. $\hat\tau(d)$ and
$\hat\sigma(d)$, for different values of the CV. Naturally, we always
have $\hat\tau(d) \leq \tau$, and in all the considered cases we
observe a steady decay of $\hat\tau(d)$ as $d$ increases, approaching
a limit value $\hat\tau(\infty)$ as $d\to\infty$, which is consistent
with the results of the limit-shape theorem \cite{Cox.81,Damron.18}.
Those limiting values are shown for both distributions in Fig.
\ref{fig:global_tau} (a), along the axis and the diagonal. We observe in all cases that $\hat\tau(\infty)$ decreases as the CV increases. For low values of the CV, the shift with respect to the mean LT (broken line) is always lower along the axis than for the diagonal. Yet, that anisotropy seems to disappear for strong disorder.\\

It is therefore relevant to consider how $\hat\tau(d)$ decays with the
distance $d$ towards their asymptotic values, $\hat\tau(\infty)$,
which is shown in Fig. \ref{fig:global_tau} (b) for the diagonal and in Fig. \ref{fig:global_tau} (c) for the axis.
In all those cases we observe a well-defined power-law
decay,

\beq
\hat\tau(d) - \hat\tau(\infty) \sim d^{-\alpha_\tau},
\label{eq:hattau_dep}
\eeq
where $\alpha_\tau\approx 2/3$ (broken line).

It is possible to provide a heuristic argument in order to understand
this behavior. The expected value of the minimum of $n$ iid random
variables extracted from a probability distribution with a bounded
support $[a,b]$ converges to $a$ --under very mild conditions-- as
$n^{-1}$, while its deviation decreases also as $n^{-1}$
\cite{David.03}. In our case, since the geodesic tends to explore a
lateral deviation of order $h\sim d^{2/3}$, it is reasonable to expect
the aforementioned scaling regime. Indeed, this is the observed
behavior for all the considered distributions.

\begin{figure}
\includegraphics[width=7cm]{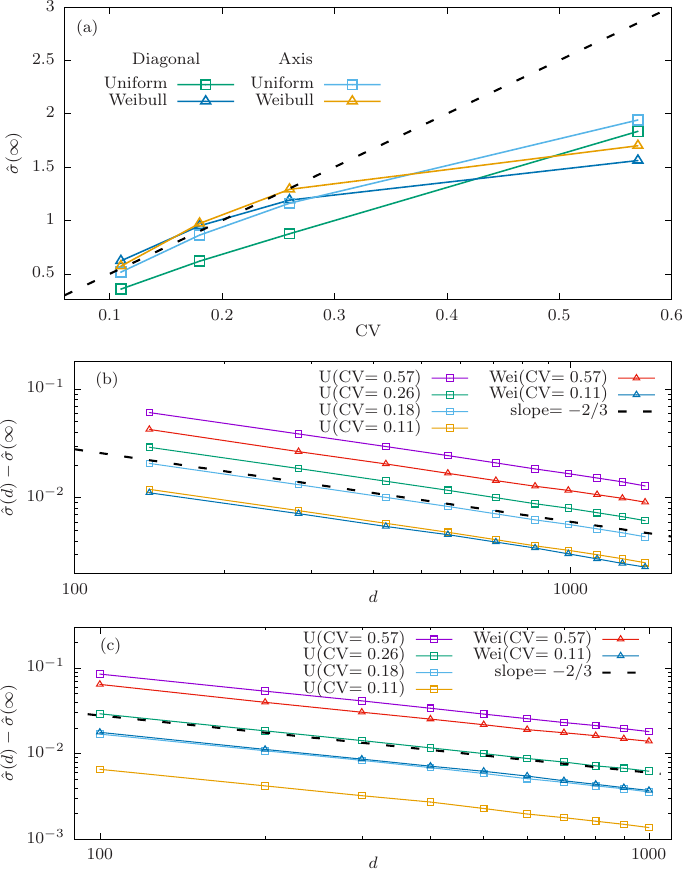}
\caption{(a) Asymptotic value of the global SLT deviation as function of the CV for the two distributions and the two lattice directions. The broken line indicates the LT deviation. (b) Convergence of the
  SLT deviations towards their asymptotic values for geodesics along
  the diagonal, as we increase the distance $d$. (c) Same data, along
  the axis. The broken line in both panels represents power-law behavior with exponent $-1/z$.}
\label{fig:global_sigma}
\end{figure}

Similarly, we may compute the global SLT deviation $\hat\sigma(d)$ as
a function of the distance $d$ for the same set of geodesics. We also
obtain a decay towards an asymptotic value $\hat\sigma(\infty)$, which
need not be lower than $\sigma$. Indeed, Fig. \ref{fig:global_sigma}
(a) shows that for low values of the CV we may have
$\hat\sigma(\infty)>\sigma$. Still, these SLT deviations converge
towards their asymptotic values following a power-law with the same
exponent as the mean,

\beq
\hat\sigma(d) - \hat\sigma(\infty) \sim d^{-\alpha_\sigma},
\eeq
with $\alpha_\sigma\approx \alpha_\tau\approx 2/3$, as we can see in
Fig. \ref{fig:global_sigma} (b) for the diagonal and Fig. \ref{fig:global_sigma} (c) for the axis.

\subsection{Distribution of the SLT sum}
\begin{figure}
\includegraphics[width=7cm]{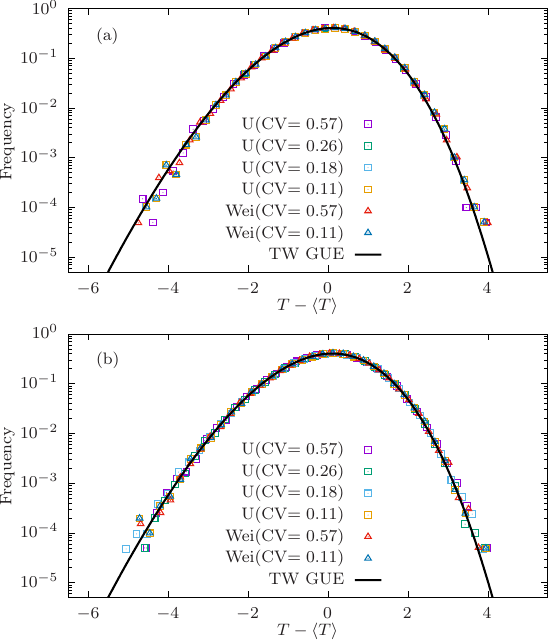}
\caption{Histograms of $\chi$, defined in Eq. \eqref{eq:fig_chi},
  which is a standardized version of the arrival time $T$, for the maximum spanning distance $d_\max$ along (a) the diagonal and (b) the axis, and different LT distributions. The suitably rescaled and
  reversed TW-GUE distribution is shown for comparison.}
\label{fig:sum_distro_TW}
\end{figure}

The SLTs are chosen to minimize the arrival time. If the SLTs were
selected through independent random sampling of the original LT
distribution, we would expect the Central Limit Theorem (CLT) to
apply. In that case, arrival time deviations
between points separated by a large distance $d$ would scale as
$\sigma_T(d)\thicksim d^{1/2}$ with normal asymptotic fluctuations.
However, we observe a scaling with the KPZ exponent $\beta=1/3$ and
values of the skewness and kurtosis that differ from those associated
with the Gaussian distribution~\cite{Cordoba.2018}.\\

In order to fully determine the nature of the arrival time
fluctuations, let us consider a standardized variable,

\beq
\chi \equiv {T - \<T\> \over \sigma_T}.
\label{eq:fig_chi}
\eeq
and construct the histograms of this variable $\chi$ for geodesics
across a distance $d_\max$ for different LT distributions. The results are
shown in Fig.~\ref{fig:sum_distro_TW} for (a) the diagonal direction
and (b) the axis. We observe an extremely good fit to a suitably
standardized TW-GUE distribution, which has been reversed along the
horizontal axis because $\chi$ exhibits negative skewness, unlike the
radial fluctuations of the isochrones \cite{Alvarez.2024}.

\subsection{Global properties of the SLT distribution}
\label{sec::KS}
\begin{figure}
\includegraphics[width=8cm]{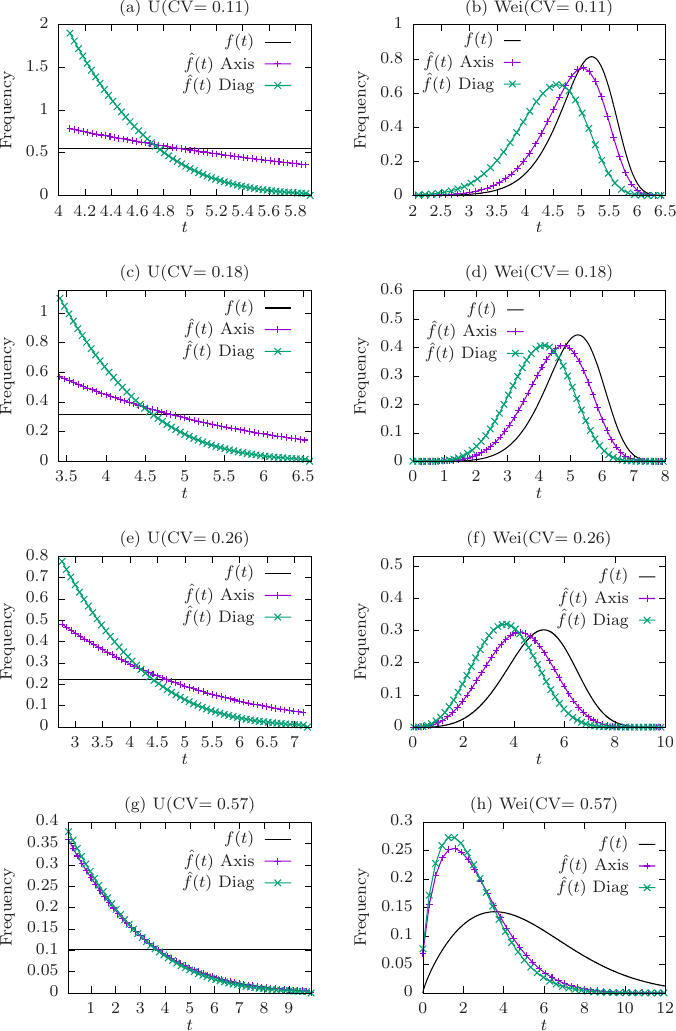}
\includegraphics[width=6cm]{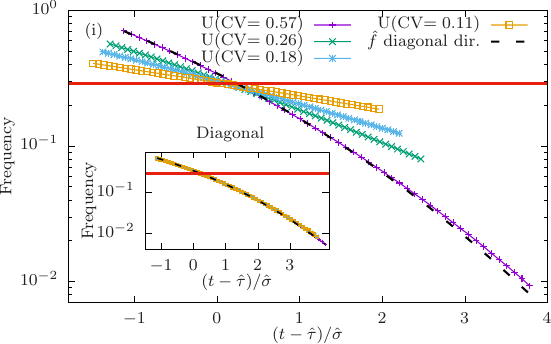}
\caption{(a)-(h) SLT histograms, $\hat f(t)$, for geodesics reaching the maximum distance $d_\max$ along the axis and the diagonal, using different LT distributions, along
  with the original LT probability density $f(t)$. Left and right
  columns show uniform and Weibull cases respectively, with CV growing
  from top to bottom. (i) Standardized SLT histograms for the uniform
  distribution along the axis and the diagonal (inset) directions, the
  horizontal line representing the original LT distribution. For the sake of comparison, we have displayed in the main figure (axis), the curve of the inset (diagonal) corresponding to CV=$0.57$.}
\label{fig:global_histograms}
\end{figure}
The mean and deviation of the selected times do not inform us about
the changes in shape between the LT and the global SLT distribution. Figures \ref{fig:global_histograms} (a) to (h) compare the corresponding histograms for geodesics spanning the maximum distance $d_\max$ both along the axis and the diagonal, and different LT distributions. On the left we observe the results for the
uniform distribution, and on the right the Weibull case, while the CV
increases from top to bottom. In each case, the corresponding original LT density funcion $f(t)$ has been displayed by a continuous line.  Notice that in all cases the SLT
histogram {\em shifts leftwards}, towards lower times. Yet, the shift
is always lower along the axis than on the diagonal, even though both
tend to become similar as the CV increases. This result agrees with the behavior of $\hat\tau(\infty)$ displayed in Fig.
\ref{fig:global_tau} (a). Figure \ref{fig:global_histograms} (i) shows all the SLT distributions for the
uniform case, after a standardization procedure, so that all of them
have zero mean and variance one. Notice that all standardized
histograms coincide approximately for the diagonal (inset), while they
converge towards the diagonal shape for the axis (main panel), as the CV
increases.\\

\begin{figure}
  \includegraphics[width=6cm]{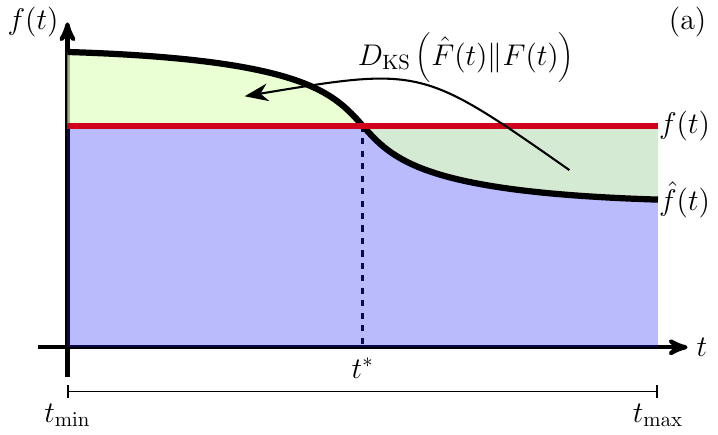}
  \includegraphics[width=6cm]{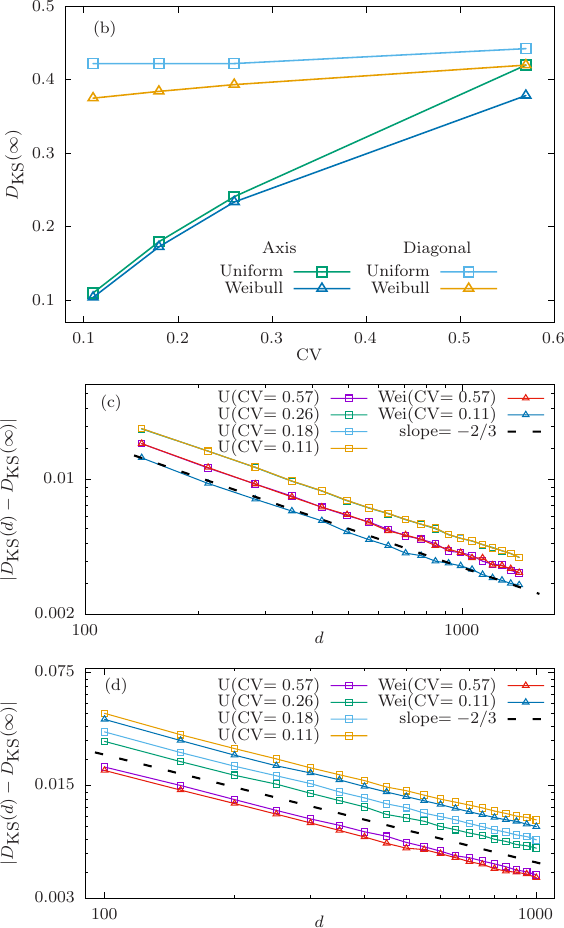}
  \caption{(a) Schematic representation of the probability shift
    between the SLT distribution (solid line) and the original LT
    distribution (horizontal red line). (b) KS distance between the
LT and the SLT distributions for asymptotically large distances along the axis and the diagonal, as a function of the CV for the two LT  distributions. (c) Convergence of the KS distances between the LT and the SLT distributions toward their asymptotic values as a function of the distance for the diagonal direction. (d) Same data, for the axis. The broken line in both panels represents power-law behavior with exponent $-1/z$.}
  \label{fig:Dks}
\end{figure}

Can we make the idea of similarity and convergence between the SLT
distributions more precise? Certainly, by making use of the {\em
  Kolmogorov-Smirnov} (KS) distance between two probability
distributions, $F(t)$ and $\hat F(t)$, defined as

\beq
D_\KS\(\hat F(t)\|F(t)\)=\sup_t \left|\hat F(t)-F(t)\right|.
\label{eq:KS}
\eeq
The KS distance between the LT and SLT distributions can be
interpreted as the degree of optimization achieved along the geodesic.
Concretely, if we have a single $t^*$ such that $f(t^*)=\hat f(t^*)$,
then the KS distance can be understood as the amount of probability
that shifts from values above $t^*$ to those below $t^*$, see Fig.
\ref{fig:Dks} (a) for an illustration.

Figure \ref{fig:Dks} (b) shows the KS distances between the LT and the
SLT distributions in the asymptotic regime, for $d\to\infty$. Notice
that, along the axis, the KS distance between both distributions grows
quickly with the CV, in a similar way for the uniform and the Weibull
distributions. Yet, along the diagonal, the asymptotic KS distance
remains nearly constant for all the levels of disorder.

It is also interesting to observe how the KS distance between the LT
and the SLT distributions approaches its asymptotic value as the
distance $d$ increases, which is shown in Fig. \ref{fig:Dks} for the
diagonal (c) and the axis (d) directions. In all cases we observe a
power-law approach,

\beq
|D_\KS(d)-D_\KS(\infty)| \sim d^{-\alpha_\KS},
\eeq
with $\alpha_\KS\approx 2/3$, as it was the case for the mean and
deviation.


\section{Local SLT distribution}
\label{sec:local}

\subsection{Local mean and deviation}

\begin{figure}
\includegraphics[width=8cm]{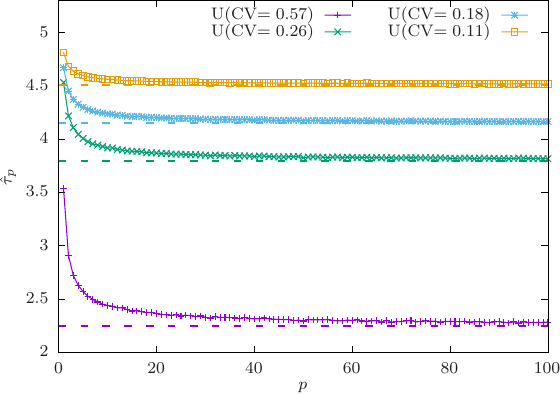}
\caption{Average value of the local SLTs, $\hat\tau_p$, as a function of the distance to the closest extreme, $p$, for geodesics reaching the maximun distance $d_\max$ along the diagonal and a uniform distribution with different CV. The horizontal lines denote the
  corresponding values of the global SLT mean, $\hat\tau$. Note that, in all cases,
  the LT mean is $\tau=5$.}
\label{fig:bulk_uni}
\end{figure}

Geodesics are specially constrained near their extremes. Therefore, it
is reasonable to consider how the SLT distribution for a given link
depends on its relative position along the geodesic. Let $p$ denote
the distance to the closest extreme, and let $\hat f_p(t)$ denote the
{\em local} SLT distribution at that position, with $\hat\tau_p$ and
$\hat\sigma_p$ standing for the corresponding mean and deviation.
Thus, we expect $\hat{\tau}_1$ to be close to $\tau$ since there are only
four options to choose the first edge. Nonetheless, as we increase
$p$, we expect $\hat\tau_p$ to decrease further, and to eventually
approach the global value $\hat\tau$. Figure \ref{fig:bulk_uni} shows
the expected value $\hat\tau_p$ as a function of $p$ for geodesics to distance $d_{\max}$ along the diagonal and different values of the CV of the uniform distribution. The horizontal lines denote the
global values $\hat\tau$, and we observe how $\hat\tau_p$ converges
towards them.

\begin{figure}
\includegraphics[width=6cm]{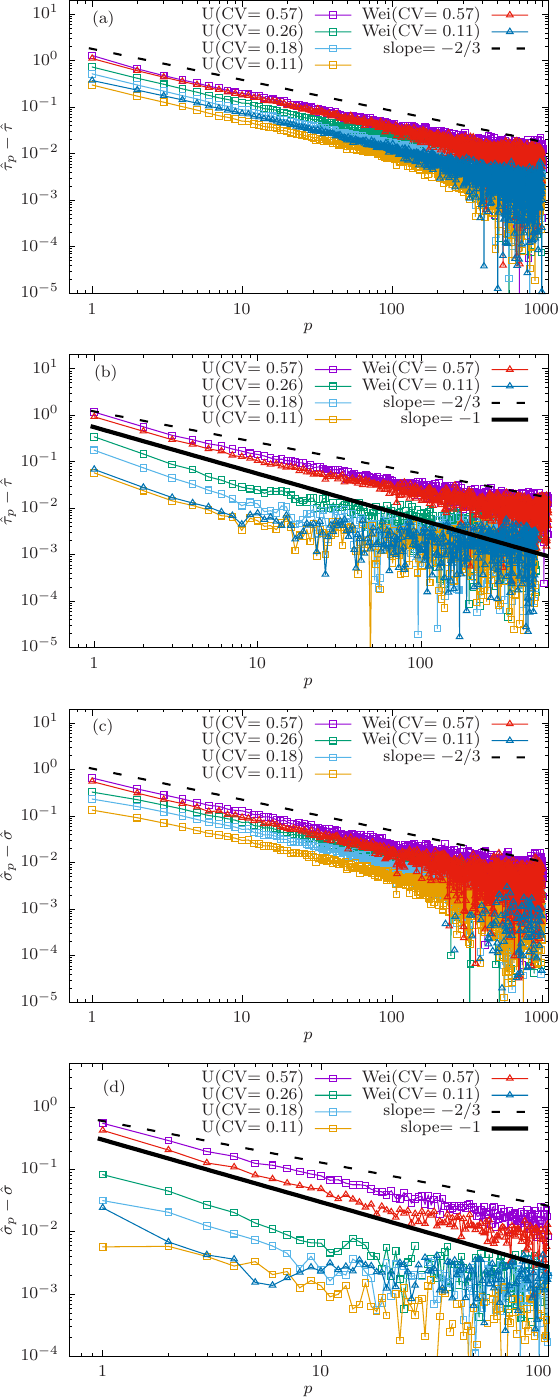}
\caption{Average local SLT excess, $\hat\tau_p-\hat\tau$, for a link at a distance $p$ from the closest extreme of geodesics reaching the maximum distance $d_\max$ along (a) the diagonal and (b) the axis, for several uniform and Weibull distributions. Local SLT excess
  deviation, $\hat\sigma_p-\hat\sigma$, for the same distributions,
  along the diagonal (c) and the axial (d) directions. Broken and solid lines represent power-law behaviors with exponents $-1/z$ and $-1$, respectively.}
\label{fig:local_mean_dev}
\end{figure}

We may now ask about the convergence of the local mean $\hat\tau_p$
towards the global mean $\hat\tau$ for different distributions and
directions, as a function of the distance $p$ to the closest extreme.
The results are shown in Fig. \ref{fig:local_mean_dev} (a) for the
diagonal and in Fig. \ref{fig:local_mean_dev} (b) for the axis, in logarithmic scale, in
order to highlight the power-law decay,

\beq
\hat\tau_p - \hat\tau \sim p^{-\alpha_p}.
\eeq
We observe that, along the diagonal, $\alpha_p\approx 2/3$ for all the
considered distributions. Indeed, this result may be accounted for
using a similar heuristic argument as that provided in Sec.
\ref{sec:global} for the decay of $\hat\tau(d)$ with the distance.
Yet, along the axial direction, the data suggest a faster decay, with
$\alpha_p\approx 1$, see Fig. \ref{fig:local_mean_dev} (b). Local
deviations, $\hat\sigma_p$, also approach their global value
$\hat\sigma$ as a power law, as we can check in Fig.
\ref{fig:local_mean_dev} (c) for the diagonal and in Fig. \ref{fig:local_mean_dev} (d) for the
axis, with similar scaling exponents. \\

Finally, it is worth noting that the constraints near the endpoints do not affect the statistics of the arrival time fluctuations. Although not shown here, we obtained the histograms of the sum of the SLTs belonging to the bulk of the geodesic, thus discarding the SLTs near the endpoints. To do this, we considered only the SLTs that satisfy $|t_p-\hat{\tau}|<\delta$, for various values of $\delta$. In all cases, we obtained the TW-GUE distribution.
\subsection{Directional effects}

\begin{figure}
\includegraphics[width=6cm]{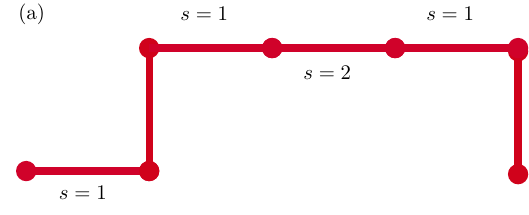}
\includegraphics[width=6cm]{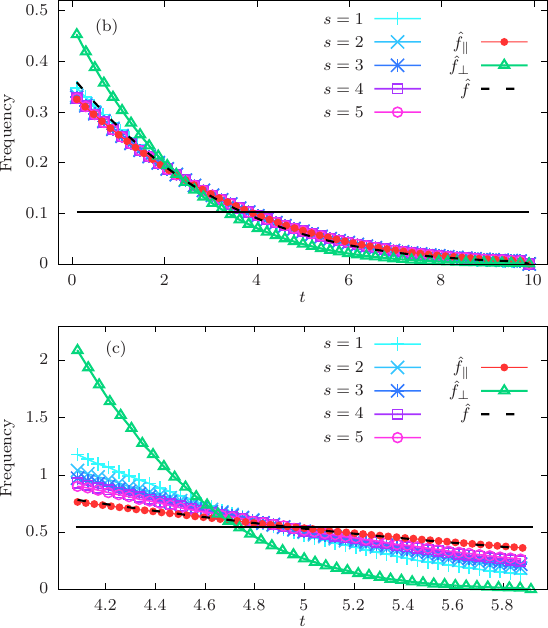}
\caption{(a) A schematic representation of the label $s$ for
  longitudinal links along an axial geodesic. (b) and (c) Histograms
  of the longitudinal and transverse SLTs, $\hat
  f_\parallel$ and $\hat f_\perp$, for a uniform disorder with CV=0.57 (b) and
  CV=0.11 (c), along with the histograms of longitudinal SLTs for different values of $s$, $\hat f_{\parallel,s}$. Corresponding global SLT histogram and original LT pdf have been displayed in both panels by the broken line and the solid black line, respectively, in order to compare. Results correspond to geodesics spanning the maximum distance $d_\max$ along the axis.}
\label{fig:link_direction}
\end{figure}

Let us consider geodesics along an axial direction. In this case we
may distinguish between {\em longitudinal} and {\em transverse links},
depending on whether they point along the axis or in the perpendicular
direction. Transverse links always increase the geodesic length and,
therefore, we expect the associated SLTs to take specially low values.
Let $\hat f_\parallel(t)$ and $\hat f_\perp(t)$ denote their respective
pdfs. Furthermore, we will label the longitudinal links according to
the distance $s$ to the closest transverse link along the geodesic, as
illustrated in Fig. \ref{fig:link_direction} (a).

Figure \ref{fig:link_direction} shows different SLT histograms for both
transverse and longitudinal selected-links for (b) a noisy uniform
distribution with CV=0.57 and (c) a low-noise uniform distribution
with CV=0.11. In each panel, we display the histograms of: all SLTs, $\hat f$, the transverse SLTs, $\hat f_{\perp}$, the longitudinal SLTs, $\hat f_{\parallel}$, and the longitudinal SLTs with different values of $s$, $\hat f_{\parallel,s}$.
Transverse SLTs always take lower values, as expected, and their
histogram is close to the global SLT histogram along the diagonal (not shown here).
Longitudinal SLTs, on the other hand, are
closer to the original LT distribution (horizontal line) in the low disorder case, Fig.
\ref{fig:link_direction} (c), and in this case they show an
interesting dependence on the distance to the closest transverse link,
$s$. Indeed, we observe that longitudinal SLTs take lower values for
$s=1$, and approach the global longitudinal histogram as $s$
increases.


\section{Correlations along the SLT}
\label{sec:correlations}
 
Once we have characterized the global properties of the SLT
distribution in Sec. \ref{sec:global}, and the local properties in
Sec. \ref{sec:local}, it is now the turn of their {\em correlation
  structure}. Despite the fact that the original LTs are iid random
variates, we are aware that selection and conditioning may induce an
intricate pattern of correlations among them, as illustrated e.g. in
Berkson's paradox \cite{Berkson.46,Pearl.18,DeRon.21}.

Theoretical arguments strongly suggest the emergence of long-range
correlations along the SLT distribution. Arrival times are sums of
SLTs, and their deviation scales as $d^\beta$, where $\beta=1/3$ is
the growth exponent of KPZ. Furthermore, the probability distribution
for the arrival times, suitably rescaled, corresponds to the
Tracy-Widom distribution for the GUE ensemble.
These two well known facts would be impossible if the SLTs were
independent random variables. Indeed, the CLT would ensure in this
case that the arrival times should be Gaussian, and their deviation
should scale as $d^{1/2}$. As we will see at the end of this section,
such a strong violation of the hypotheses of the CLT imposes very
strict conditions on the correlations among the SLTs. 

\subsection{Two-point correlations}

Let us define $\bar t_p\equiv\hat t_p - \hat\tau_p$, so that the
covariance between SLTs located at positions $p$ and $q$ along the
geodesic becomes

\beq
\text{Cov}(p,q) \equiv \< \bar t_p\, \bar t_q \>.
\eeq
The (two-point) correlation between them is then defined as

\beq
C(p,q) \equiv {\text{Cov}(p,q) \over \hat\sigma_p \hat\sigma_q}.
\eeq
Furthermore, if we expect a translationally invariant structure we may
define the correlation function as a spatial average,

\beq
C(k) \equiv \overline{C(p,p+k)}.
\eeq
Naturally, we are aware that the SLTs do not constitute a
translationally invariant random process. Yet, a {\em bulk} regime is
achieved in the central part of large geodesics.

\begin{figure}
\includegraphics[width=8cm]{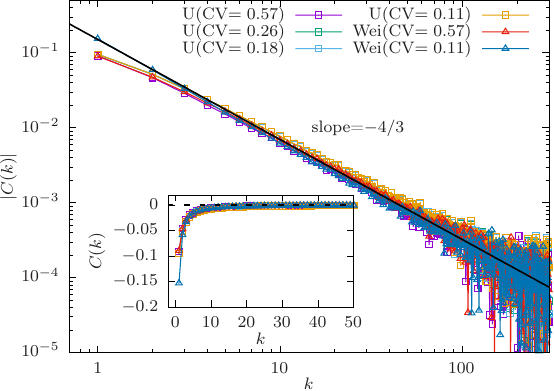}
\caption{Correlation function $C(k)$ for geodesics reaching the maximun distance $d_\max$ along the diagonal and different disorder distributions. The main panel shows the log-log plot of the absolute value, while the inset displays the bare values in linear scale. The solid line represents power-law behavior with exponent $-4/3$.}
\label{fig:corrdiag}
\end{figure}

Figure \ref{fig:corrdiag} shows the the absolute value of the correlation funcion $C(k)$ for geodesics spanning the maximum distance $d_{\max}$ along the diagonal, and different disorder distributions. The
correlation function $C(k)$ is always {\em negative} in this case (see inset),
because selecting a specially low link-time may force us to pay some
price in the neighboring link times, which tend to be slightly larger than
their mean. We also observe that correlations decay as a
power-law with a common scaling exponent,

\beq
C(k) \sim k^{-\gamma},
\eeq
with $\gamma=4/3$, which we can associate to the KPZ universality
class. Indeed, the variance of the arrival times can be computed as

\beq
\sigma^2_T =\<T^2\>-\<T\>^2 = \sum_{p,q} C(p,q) \thicksim d^{2-\gamma},
\eeq
where we have made use of the fact that $\ell \propto d$
in the weak disorder regime. Since KPZ scaling demands that
$\sigma^2_T \sim d^{2\beta}$, we deduce that $\gamma=2-2\beta=4/3$. In
other words: {\em the correlation scaling exponent along the SLTs can
  be readily linked to the KPZ class.}

\begin{figure}
  \includegraphics[width=7cm]{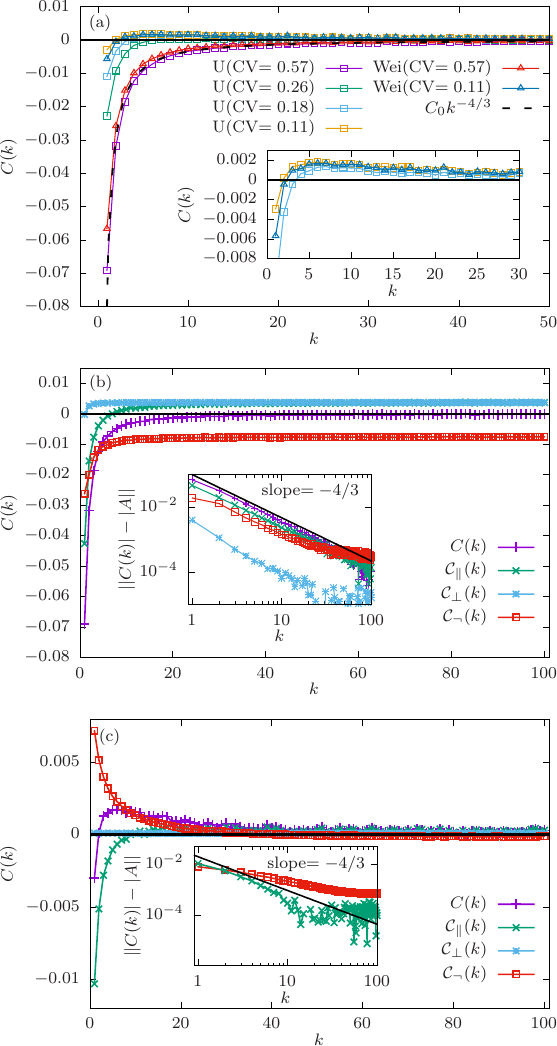}
  \caption{(a) Correlation function, $C(k)$, for geodesics reaching the maximun distance $d_\max$ along the axis and different
disorder distributions. A power law with exponent $-4/3$, obtained by fitting the correlation function for the higher-disorder case (CV=0.57), is also shown. The inset shows a zoom of the positive values. (b)-(c) Correlation function, $C(k)$, and its directional contributions, $\mathcal{C}_{\parallel}(k)$, $\mathcal{C}_{\perp}(k)$ and $\mathcal{C}_{\neg}(k)$, for geodesics reaching the maximum distance $d_\max$ along the axis and a uniform distribution with (b) CV=0.57 and (c) CV=0.11. Insets show their convergence towards their asymptotic values, represented by $A$. }
\label{fig:corraxis}
\end{figure}

Figure \ref{fig:corraxis} (a) shows the correlation function, $C(k)$, along the axis, for several disorder distributions and for geodesics reaching the maximum distance $d_{\max}$. For the higher disorder case (CV = 0.57), we observe that the correlation function $C(k)$ tends to zero as a power law with the same exponent obtained along the diagonal. However, the sign of the correlation function behaves in a surprising way for low disorder (CV = 0.11), since $C(k)<0$ for very small $k$, but then jumps to a large positive value, from which it decays to zero.\\

In order to understand the complex pattern displayed by the correlation function along the axis, we should consider their {\em directional
  information}. In Sec. \ref{sec:local} we distinguished between
longitudinal and transverse links, and we may make a similar
distinction here. Given a geodesic, let us consider the terms of the
form $\pi(p,q)\equiv\bar t_p \bar t_q /\hat\sigma_p \hat\sigma_q$
which involve different combinations of link types. We may distinguish
three such combinations:

\begin{itemize}
  \item Longitudinal contribution, $\pi_\parallel(p,q)$, if both $p$ and $q$
  are longitudinal links.
  \item Transverse contribution, $\pi_\perp(p,q)$, if both $p$ and $q$ are
  transverse links.
  \item Mixed contribution, $\pi_\neg(p,q)$, if one of them is
  longitudinal and the other is transverse.
\end{itemize}
Then, we perform the average over all products of the same type within each geodesic as a function of $k=p-q$, to obtain the following decomposition of the correlation function:
\begin{equation}
\begin{aligned}
C(k)=&\overbrace{\<\overline{\pi_\parallel (p,p+k)}\>}^{\mathcal{C}_{\parallel}(k)}+\overbrace{\<\overline{\pi_\perp (p,p+k)}\>}^{\mathcal{C}_{\perp}(k)}\\
&+\underbrace{\<\overline{\pi_\neg (p,p+k)}\>}_{\mathcal{C}_{\neg}(k)},
\end{aligned}
\end{equation}
where the overline indicates the appropriate average over products of
the same type within each geodesic.

Let us now return to Fig. \ref{fig:corraxis}, and observe the behavior
of the directional contributions to the correlation functions. For
high CV, see Fig. \ref{fig:corraxis} (b), all three contributions grow monotonously. The
longitudinal term $\mathcal{C}_\parallel(k)$ starts out negative for
very low $k$, but finally converges to a positive value for large $k$.
The transverse term, $\mathcal{C}_\perp(k)$, is always positive, and
the mixed term remains negative. The inset shows their convergence
towards their asymptotic value, and we can see that their behaviors
are compatible with the global expected scaling. Noticeably, the three
contributions approach asymptotically a nonzero value. To undestand
this behavior, let us recall that the average of the SLTs
at positions $p$ and $q$ are obtained without taking into account
the link direction at those positions. If we denote by
$\hat{\tau}_{\perp}$ and $\hat{\tau}_{\parallel}$ the mean values of
the transverse and longitudinal SLTs, respectively, we can deduce from Fig. \ref{fig:link_direction} the following inequalities:
\begin{equation}
\hat{\tau}_{\perp}\leq \hat{\tau}\leq\hat{\tau}_{\parallel},
\label{eq:inequalities mean values}
\end{equation}
where equality holds in the case
$\hat{f}_{\perp}=\hat{f}_{\parallel}$. Indeed, the sign of the asymptotic values of each contribution is primarily determined by the inequalities in Eq.~\eqref{eq:inequalities mean values}. In the case of the transverse contribution, $\mathcal{C}_{\perp}(k)$, we have $\hat{\tau}_{\perp}-\hat{\tau}<0$, and the sign of its asymptotic value is given by the sign of the product $(\hat{\tau}_{\perp}-\hat{\tau})(\hat{\tau}_{\perp}-\hat{\tau})>0$. On the other hand, for the longitudinal contribution, $\mathcal{C}_{\parallel}(k)$,  we have $\hat{\tau}_{\parallel}-\hat{\tau}>0$, and thus $(\hat{\tau}_{\parallel}-\hat{\tau})(\hat{\tau}_{\parallel}-\hat{\tau})>0$. For the mixed contribution, $\mathcal{C}_{\neg}(k)$, the cross product $(\hat{\tau}_{\parallel}-\hat{\tau})(\hat{\tau}_{\perp}-\hat{\tau})$ is negative, thus explaining the sign of all contributions.\\

Finally, Fig. \ref{fig:corraxis} (c) shows the same data for a low disorder
uniform distribution with CV=0.11. The longitudinal term is positive and grows monotonously, as in the high-disorder case displayed in Fig. \ref{fig:corraxis} (b), and the transverse term is almost negligible. Remarkably, the mixed contribution is now positive and decays monotonously. These behaviors provide a route to explain the complex behavior
of the final correlator. To understand these patterns, we need to draw attention to the fact that, for weak disorder, the number of transverse SLTs is much smaller than the number of longitudinal SLTs, and two consecutive
transverse SLTs constitute an event of negligible probability.
Consequently, the value of the local mean $\hat{\tau}_p$ is close to
the mean associated with longitudinal SLTs, $\hat{\tau}_{\parallel}$,
resulting in a small positive asymptotic value for
$\mathcal{C}_{\parallel}$. In addition, the relatively small number of transverse links explains the low value of the transverse contribution $\mathcal{C}_{\perp}$. To explain the behavior of the mixed contribution, $\mathcal{C}_{\neg}(k)$, we recall the results displayed in Fig. \ref{fig:link_direction} (b), which showed that longitudinal links tend to have lower values as they approach the closest transverse link. If we denote by $\hat{\tau}_{\parallel,s}$ the mean value of the longitudinal SLTs at a distance $s$ from the closest transverse link, we have
that the mixed contribution is governed by terms of the form $(\hat{\tau}_{\parallel, k}- \hat\tau)(\hat\tau_{\perp}-\hat\tau)$, which are positive por small values of $k$. As $k$ increases, $\hat{\tau}_{\parallel, k}$ approaches $\hat\tau$, thus yielding the monotonically decreasing behavior of the contribution $\mathcal{C}_{\neg}(k)$.

\subsection{High-order correlations and the violation of the CLT hypotheses}

Yet, long-range two-point correlations need not lead to a violation of
the hypotheses of the CLT
\cite{Feller.68,Ibragimov.71,Billingsley.95}, which in this case is
ruled by the Breuer-Major theorem \cite{Peccati.11,Pipiras.17}. In
intuitive terms, if the two-point correlation fulfills some mild
conditions, we may perform a Cholesky decomposition of the covariance
matrix, and express the SLT vector as a linear transformation of
another vector of iid random variates, for which the CLT holds.
Therefore, our previous results do not explain by themselves the
emergence of a strongly non-Gaussian limit distribution for the SLT
sums, i.e. the Tracy-Widom distribution, which is not even infinitely
divisible \cite{Dominguez.16}.

Indeed, in order to ensure Gaussianity of the sum of $n$ correlated
variables $\{X_i\}$ with zero average and finite covariance
$K(X_i,X_j)$, the higher-order correlations must fulfill Wick's
theorem (also known as Isserlis' theorem)
\cite{Peccati.11,Pipiras.17},

\beq
\<X_1\cdots X_n\> = \begin{cases}
  0, & \text{$n$ odd,}\\
  \sum_{\text{Pairings}} \prod_{(i,j)}
  K(X_i,X_j), & \text{$n$ even.}
\end{cases}
\eeq

Since the sum of the SLTs is strongly non-Gaussian, the higher-order
correlations must deviate from the Wick prescription. This is a usual
phenomenon observed in field theory and condensed matter physics, in
which the integral (or sum) of a stochastic or quantum field over a
finite region is shown to deviate from Gaussianity due to the presence
of non-Wick higher-order correlations
\cite{Fewster.18,Camia.16,Anthony.19}, which is known as {\em
  full-counting statistics}.

Specifically, the conformally invariant case presents a broad interest
due to its applicability to critical phenomena \cite{DiFrancesco.96}.
For a 1+1D conformal field theory (CFT), the form of the two-point and
the three-point correlators is well known,

\begin{align}
  \<X_iX_j\> &\approx {A_2 \over |i-j|^{2\Delta}}, \\
  \<X_iX_jX_k\> &\approx {A_3 \over |i-j|^\Delta |i-k|^\Delta
    |j-k|^\Delta},
\end{align}
where $\Delta$ is the conformal weight associated to the field, and
$A_2$ and $A_3$ are constants.

\medskip

\begin{figure}
\includegraphics[width=8cm]{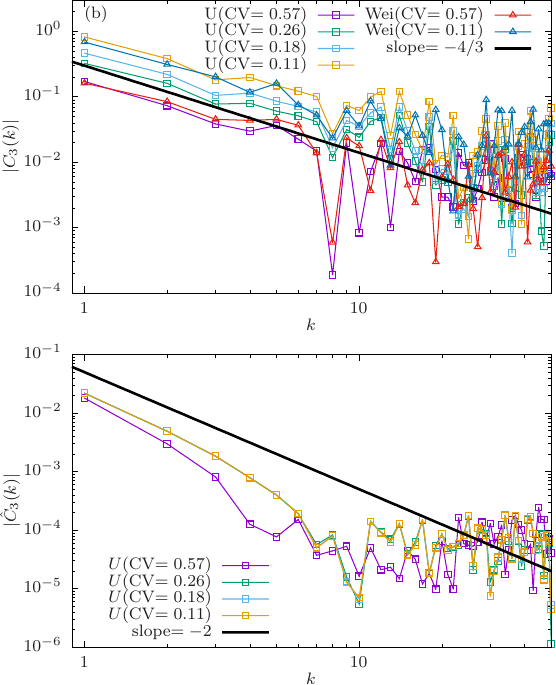}
\caption{(a) Three-point correlation function $C_3(k)$, defined in Eq. \eqref{eq:c3}, for geodesics reaching $d_\max$ along the diagonal and different uniform and Weibull distributions. The solid line represents the predicted $k^{-4/3}$ decay. (b) Alternative three-point correlation function $\hat{C}_3(k)$ defined in Eq. \eqref{eq:c3prime}, and the theoretically
  expected $k^{-2}$ decay.}
\label{fig:corr3}
\end{figure}

In our case, non-Gaussianity suggests that the three-point correlators
should be non-zero. The generic power-law behavior observed within the
KPZ class allows us to attempt a CFT Ansatz, even though we do not
claim at this stage any conformal invariance in our model. Compliance
with the two-point function requires that we use $\Delta=2/3$. Thus,
we hypothesize

\beq
\<\bar t_i \bar t_j \bar t_k\> \propto {1\over |i-j|^{2/3} |i-k|^{2/3}
  |j-k|^{2/3}}.
  \label{eq:prediction c3(k)}
\eeq
Yet, we can perform a sanity check of this expression before advancing
further. Does it yield a finite skewness for the arrival times, $\bar
T = T - \<T\>$? According to our Ansatz, the third moment scales as

\beq
\<{\bar T}^3\> = \sum_{ijk} \<\bar t_i \bar t_j \bar t_k\> \thicksim
d^{3-3\Delta}.
\eeq
Indeed, from the definition of the skewness we obtain
\beq
\gamma_{\bar T} = {\<{\bar T}^3\> \over \<{\bar T}^2\>^{3/2} }
\sim  {d^{3-3\Delta} \over \(d^{2-2\Delta}\)^{3/2}} =
\text{Const,}
\eeq
so, the Ansatz makes sense. Does it actually work? We show in Fig. \ref{fig:corr3} the results for two different three-point correlators, for geodesics spanning distance $d_{\max}$ along the diagonal and different LT distributions. Figure \ref{fig:corr3} (a) shows the decay of $C_3(k) \equiv \<\bar t_i \bar t_{i+1} \bar t_{i+k}\>$ and we check that it agrees with the prediction from Eq. \eqref{eq:prediction c3(k)}, 
\beq
C_3(k) \equiv \<\bar t_i \bar t_{i+1} \bar t_{i+k}\> \sim k^{-2\Delta}
= k^{-4/3}.
\label{eq:c3}
\eeq
Figure \ref{fig:corr3} (b) shows the decay of $\hat{C}_3(k)\equiv \<\bar t_i \bar t_{i+k} \bar t_{i+2k} \>$, for which we expect
\beq
\hat C_3(k) \equiv \<\bar t_i \bar t_{i+k} \bar t_{i+2k} \> \sim
k^{-3\Delta} = k^{-2}.
\label{eq:c3prime}
\eeq
The faster decay for this type of correlators is harder to capture
numerically, yet Fig. \ref{fig:corr3} (b) manages to catch a glimpse
for low values of $k$.

\section{Conclusions and further work}

In this work, we have explored the statistical properties of the SLTs,
i.e. the link-times {\em actually selected to form part of different
  geodesics} in FPP lattices. Naturally, they tend to take lower
values than those prescribed by the general LT distribution. Moreover,
in similarity to Berkson's paradox, we observe emergent correlations
among them induced by the selection procedure, despite the fact that
the original LTs are iid random variables.

Specifically, we have considered geodesics --also known as optimal
paths-- joining points along an axis or along a diagonal of a square
lattice, using probability distributions for the LTs in the uniform
and Weibull families, for different values of the CV. The SLT
probability distribution displays scaling behavior in many regimes.
For example, the global mean and deviation of the SLT distribution
decay as the distance is increased, approaching a limiting value as a
power law with exponent $-2/3$, which we have associated to KPZ
behavior. Moreover, geodesics are typically more constrained near
their extreme points, and the mean and deviation of the local SLT distribution approach the bulk behavior in a similar fashion.
Intriguing directional effects are observed when the geodesic extends
along an axis, which require further investigation. Finally, the
selection mechanism induces long-range correlations among the SLTs, which decay with their distance along the geodesic
as a power-law with exponent $-\gamma=2\beta-2=-4/3$, where
$\beta=1/3$ is the growth exponent of KPZ.

The arrival times in FPP, which are the sums of SLTs, converge to a
Tracy-Widom distribution. Therefore the SLTs must violate some of the
hypotheses underneath the CLT. We have explored the origin of such
violations, and found non-Wick long-range correlations of
third order. An Ansatz inspired in conformal symmetry arguments has
allowed us to conjecture a scaling form for the three-point
correlation function.

The statistical properties of the SLTs have proved to be far richer
than initially expected, and many features deserve further research.
For example, in this work we have neglected the effects of the total
length of the geodesic. Elucidation of the origin of the TW as a sum
of identically distributed but highly correlated SLTs is still
incomplete. Indeed, sums of identically distributed but strongly
correlated random variables may still be Gaussian if the covariance
matrix follows certain regularity constraints, as described in the
Breuer-Major theorem, and fall into the different Hermite processes
when a non-linear function acts on the variables
\cite{Peccati.11,Pipiras.17}. Still, this framework does not seem to
provide a route to show how the TW distribution may emerge in this
context. This question poses an intriguing problem for the
mathematical physics community.

\begin{acknowledgments}
  This work has benefited from discussions with R. Cuerno, J.M. López
  and D. Villarrubia. We acknowledge the Spanish government for
  financial support through grants PID2021-123969NB-I00,
  PID2024-159024NB-C21 and PID2024-159024NB-C22, as well as the
  computational resources and assistance provided by the Centro de
  Computación de Alto Rendimiento (CCAR-UNED). I.A.D. acknowledges
  funding from UNED through an FPI scholarship.
\end{acknowledgments}


\end{document}